\documentclass[draftcls,onecolumn,12pt]{IEEEtran}%
\usepackage{epsfig,makeidx,color}
\usepackage{graphicx}
\usepackage{amsbsy}
\usepackage{amsmath}
\usepackage{amssymb}
\usepackage{euscript}
\usepackage{citesort}
\usepackage{fancyhdr}

\newtheorem{mylemma}{\bf Lemma}

\def\beq{\begin{equation}}
\def\eeq{\end{equation}}

\begin{document}

\title{On the Performance of \\ Golden Space-Time Trellis Coded Modulation \\ over MIMO Block Fading Channels}

\author{Emanuele~Viterbo and Yi~Hong
\thanks{\scriptsize Emanuele~Viterbo is with DEIS -
Universit\`{a} della Calabria, via P.~Bucci, 42/C, 87036 Rende (CS),
Italy, e-mail: $\tt viterbo$@$\tt deis.unical.it$. Yi~Hong was DEIS
- Universit\`{a} della Calabria, Italy, and is now with Institute of
Advanced Telecom., University of Wales, Swansea, SA2 8PP, UK,
e-mail: $\tt y.hong$@$\tt swansea.ac.uk$.}}

\maketitle
%%%%%%%%%%%%%%%%%%%%%%%%%%%%%%%%%%%%%%%%%%%%%%%%%%%%%%%%%%%%%%%%%%%%%%
\begin{abstract}
The Golden space-time trellis coded modulation (GST-TCM) scheme was
proposed in \cite{Hong06} for a high rate $2\times 2$ multiple-input
multiple-output (MIMO) system over slow fading channels. In this
letter, we present the performance analysis of GST-TCM over block
fading channels, where the channel matrix is constant over a
fraction of the codeword length and varies from one fraction to
another, independently. In practice, it is not useful to design such
codes for specific block fading channel parameters and a robust
solution is preferable.
%We therefore analyze the performance of
%GST-TCM designed for slow fading over arbitrary block fading channels.
%The impact of the block fading channel on the code
%performance is analyzed based on the design criteria of GST-TCM.
We then show both analytically and by simulation that the GST-TCM
designed for slow fading channels are indeed robust to all block
fading channel conditions.
\end{abstract}

%%%%%%%%%%%%%%%%%%%%%%%%%%%%%%%%%%%%%%%%%%%%%%%%%%%%%%%%%%%%%%%%%%%%%%
\section{Introduction}

The Golden code was proposed in \cite{Golden05} as a full rate and
full diversity code for $2\times 2$ multiple-input multiple-output
(MIMO) systems with {\em non-vanishing minimum determinant} (NVD).
It was shown in \cite{Elia05} how this property guarantees to
achieve the diversity-multiplexing gain trade-off.
In order to enhance the coding gain, a first attempt to concatenate
the Golden code with an outer trellis code was made in
\cite{David05}. However, the resulting {\em ad hoc} scheme suffered
from a high trellis complexity.

In \cite{Hong06}, a Golden space-time trellis coded modulation
(GST-TCM) scheme was designed for slow fading channels. The NVD
property of the inner Golden code is essential for a TCM scheme.
This property guarantees that the code will not suffer from a
reduction of the minimum determinant, when a constellation expansion
is required \cite{Golden05}. The systematic design proposed in
\cite{Hong06}, is based on set partitioning of the Golden code in
order to increase the minimum determinant. An outer trellis code is
then used to increase the Hamming distance between the codewords.
The Viterbi algorithm is applied for trellis decoding, where the
branch metrics are computed with a lattice {\em sphere decoder}
\cite{VB1,Viterbo99} for the inner Golden code.

In this letter, we analyze performance of the GST-TCM scheme in {\em
block fading} channels \cite{Biglieri98}. The block fading channel
is a simple and powerful model to describe a variety of wireless
fading channels ranging from fast to slow. For example, in OFDM
based systems over frequency selective fading channels it can model
various channel delay profiles. In particular, low delay spread
channels correspond to small frequency selectivity, i.e., many
adjacent subcarriers experience similar fading coefficients. On the
contrary, channels with long delays profiles correspond to large
frequency selectivity, i.e., the fading coefficients vary
significantly among adjacent subcarriers.

In practice, it is not useful to design a GST-TCM for specific block
fading channel parameters and a robust solution is preferable. We
therefore analyze the performance of known GST-TCM, designed for
slow fading, over arbitrary block fading channels. The impact of the
block fading channel on the code performance is estimated
analytically using a two-term truncated union bound (UB). We finally
show both analytically and by simulation that the GST-TCM designed
for slow fading channels are indeed robust to various block fading
channel conditions.

The rest of the letter is organized as follows.
Section~\ref{sec:sysmod} introduces the system model for block
fading channels. Section~\ref{sec:designcriteria} presents an
analytic performance estimation of linear STBCs over block fading
channels. In Section~\ref{sec:performance} we specialize the result
for GST-TCM designed for slow fading. Section~\ref{simuandresults}
shows simulation results. Conclusions are drawn in Section VI.

{\em Notations}: Let $T$ denote transpose and $\dagger$ denote
Hermitian transpose. Let $\mathbb{Z}$, $\mathbb{C}$ and
$\mathbb{Z}[i]$ denote the ring of rational integers, the field of
complex numbers, and the ring of Gaussian integers, respectively,
where $i^2 =-1$. Let $\lceil x \rceil$ denote the smallest integer
greater or equal to $x$. The operator $\bar{(\cdot)}$ denotes the
algebraic conjugation in a quadratic algebraic number field
\cite{Golden05}.
%%%%%%%%%%%%%%%%%%%%%%%%%%%%%%%%%%%%%%%%%%%%%%%%%%%%%%%%%%%%%%%%%%%%%%
%%%%%%%%%%%%%%%%%%%%%%%%%%%%%%%%%%%%%%%%%%%%%%%%%%%%%%%%%%%%%%%%%%%%%%
\section{System Model}\label{sec:sysmod}

Let us first consider a $2\times 2$ MIMO system ($n_T = 2$ transmit
and $n_R = 2$ receive antennas) over a slow fading channel using the
Golden code $\mathcal{G}$. A $2\times 2$ Golden codeword $X \in
\mathcal{G}$ is transmitted over two channel uses, where the channel
matrix $H$ is constant and
\begin{equation}\label{mimo2x2system}
Y = H X + Z
\end{equation}
is received, where $Z$ is a complex white Gaussian noise $2\times 2$
matrix. The Golden codeword $X \in \mathcal{G}$ is defined as
\cite{Golden05}
\begin{equation}
    X\triangleq\frac{1}{\sqrt{5}}\left[
    \begin{array}{cc}
    \alpha \left( a+b\theta \right)  & \alpha \left( c+d\theta \right)\\
    i \bar{\alpha} \left( c+d\bar{\theta} \right)  & \bar{\alpha}
    \left(  a+b\bar{\theta} \right)
    \end{array}%
    \right]  \label{goldencodeword}
\end{equation}
where $a,b,c,d \in \mathbb{Z}[i]$ are the information symbols,
$\theta \triangleq 1- \bar{\theta } =  \frac{1+\sqrt{5}}{2}$,
$\alpha \triangleq 1 + i\bar\theta$, $\bar{\alpha} \triangleq
1+i\theta$, and the factor ${1}/{\sqrt{5}}$ is used to normalize
energy \cite{Golden05}. As information symbols, $Q$-QAM
constellations are used, where $Q = 2^{\eta}$. The QAM constellation
is assumed to be scaled to match $\mathbb{Z}[i]+(1+i)/2$.

In this letter we will consider linear codes of length $L$ over an
alphabet $\mathcal{G}$ in a block fading channel, i.e., the
transmitted codewords are given by ${\bf X} = (X_1,\ldots,
X_t,\ldots, X_L)\in \mathbb{C}^{2\times 2L}$:
\begin{itemize}
\item if the elements $X_t \in \mathcal{G}$ are selected independently, we have
the {\em uncoded Golden code};
\item if a trellis outer code is used to constrain the $X_t$'s,
we have a GST-TCM \cite{Hong06}.
\end{itemize}

Let ${\bf Z} = (Z_1, \ldots, Z_t, \ldots, Z_L) \in
\mathbb{C}^{2\times 2L}$ denote a complex white Gaussian noise
matrix with i.i.d. samples distributed as $\mathcal{N}_{\mathbb{C}}
(0,N_0)$, where $Z_t$ are the complex white Gaussian noise $2\times
2$ matrices. At the receiver, we have the following received signal
matrix
\[
 {\bf Y} = (Y_1, \ldots, Y_t, \ldots,
Y_L) \in \mathbb{C}^{2\times 2L}
\]
where $Y_t$ is given by
\begin{equation}\label{systemmodel}
Y_t = H_t X_t + Z_t \hspace{1cm} t=1,\ldots, L
\end{equation}
where $H_t$ are assumed to be i.i.d. circularly symmetric Gaussian
random variables $\sim \mathcal{N}_{\mathbb{C}} (0,1)$.

In a {\em block fading channel}, the matrices $H_t \in
\mathbb{C}^{2\times 2}$ are assumed to be constant in a block of $N$
consecutive alphabet symbols in $\mathcal{G}$ (i.e., $2N$ channel
uses) and vary independently from one block to another, i.e.,
\[
H_{kN+1}=\cdots = H_{(k+1)N} ~~~\mbox{for}~~~ k=0,\ldots,L/N-1
\]
where we assume for convenience that $N$ divides $L$. This implies
that the number of blocks within a codeword experiencing independent
fading is $B = L/N$. For $N=L$ ($B=1$) we have a {\em slow} fading
channel and for $N=1$ ($B=L$) a {\em fast} fading channel. In this
letter, we assume that the channel is known at the receiver but not
at the transmitter.

%%%%%%%%%%%%%%%%%%%%%%%%%%%%%%%%%%%%%%%%%%%%%%%%%%%%%%%%%%%%%%%%%%%%%%%
\section{Performance of linear STBC over block fading channels}
\label{sec:designcriteria}

In this section we analyze performance of linear STBC over block
fading channels. In the following we will make the analysis specific
to the GST-TCM.

Assuming that a codeword $\mathbf{X}$ is transmitted over a {\em
slow fading} channel ($N=L$), the maximum-likelihood receiver might
decide erroneously in favor of another codeword $\hat{\mathbf{X}}$,
resulting in a {\em pairwise error event}. Let $r$ denote the rank
of the {\em codeword difference matrix}
$\mathbf{X}-\hat{\mathbf{X}}$. Let $\lambda_{j}, j=1,\ldots, r$, be
the non-zero eigenvalues of the {\em codeword distance matrix}
$\mathbf{A}= (\mathbf{X} -\hat{\mathbf{X}})(\mathbf{X}
-\hat{\mathbf{X}})^{\dagger}$. The {\em pairwise error probability}
(PEP) depends on the determinant $\det({\mathbf{A}})$ for full rank
codes ($r=2$) \cite{Tarokh98}.

The UB gives an upper bound to the performance of the STBC, while a
truncated UB gives an asymptotic approximation \cite{Biglieri05}.
The dominant term in the UB is the PEP that depends on the {\em
minimum determinant} of the codeword distance matrix
\[
\Delta^{(s)}_{\min}=\underset{\mathbf{X} \neq \hat{\mathbf{X}}}
{\min}\det\left( \mathbf{A} \right)
\]
where the superscript $s$ denotes the slow fading case. The
traditional code design criterion for space-time codes in
\cite{Tarokh98} is based on the minimization of the dominant term in
the UB, which in turn depends on  the {\em diversity gain} $n_{T}
n_{R}$ and the {\em coding gain} $\left( \Delta^{(s)}_{\min}
\right)^{\frac{1}{n_{T}}}$.

In this letter, we will consider the truncated UB with two terms
\begin{equation} \label{eq:truncUB}
P(e) \approx N_{s_1} P_1 +  N_{s_2} P_2
\end{equation}
where the $P_{i}, i=1,2,$ are the two largest PEPs of the two
dominating events depending on and $N_{s_i}$ the corresponding
multiplicities. We assume that $P_1$ depends on $\Delta_1
=\Delta^{(s)}_{\min}$ and $P_2$ depends on $\Delta_2$ the second
smallest value of $\det\left( \mathbf{A} \right)$.

Since we focus on {\em full rank} (i.e., $r=n_{T}=2$ for all
$\mathbf{A}$) and {\em linear} (i.e., the sum of any two codewords
is a codeword) codes, we can simply consider the PEP from the
all-zero transmitted codeword matrix.
% by using $\mathbf{A}= \mathbf{X} \mathbf{X}^\dagger$.

Let us now consider a {\em block fading} channel, where $H_t$ is
constant for $2N$ channel uses and changes independently  in the  $B
= L/N$ blocks. For a given codeword $\mathbf{X}$, we define the
matrices
\begin{eqnarray}\label{FL}
F_{\ell} &\triangleq& \sum_{t=(\ell -1)N+1}^{\ell
N}X_{t}X_{t}^{\dagger}~~~~~~~\ell=1, \ldots, B
\end{eqnarray}
Following \cite{Tarokh98}, it can be easily shown that the
dominanting term in the UB will be driven by the quantity
\begin{eqnarray}\label{Determinant}
\Delta^{(b)}_{\min } &\triangleq&
 \min_{\det(F_{\ell})\neq 0} {\prod_{\ell =1}^{B}} \det(F_{\ell})
\end{eqnarray}
where the superscript $b$ denotes the block fading case. The above
performance metric $\Delta^{(b)}_{\min}$ could hard to exploit, due
to the non-additive nature of the determinant metric in
(\ref{Determinant}). Since $\mathit{X}_{t} \mathit{X}_{t}^{\dagger}$
are positive definite matrices, we resort to the following
determinant inequality \cite{matrixbook}
\begin{eqnarray}\label{DetFL}
\det (F_{\ell}) \geq {\sum_{t=(\ell -1)N+1}^{\ell N}} \!\!\!\!\!\det
\left(X_{t}X_{t}^{\dagger }\right) ~\triangleq~ a_{\ell}
\end{eqnarray}
and use the simpler lower bound:
\begin{eqnarray}\label{Detmin}
\Delta^{(b)}_{\min } \geq \underset{{a_{\ell} \neq 0}} \min~
{\prod_{\ell=1}^{B}}~ a_{\ell} \triangleq \Delta^{(b)'}_{\min }
\end{eqnarray}
We can see that the $\Delta^{(b)'}_{\min }$ is not only determined
by the code structure, but also by the block fading channel
parameters $B$ and $N$. Note that $\Delta^{(b)'}_{\min}$ coincides
with the $\Delta'_{\min}$ defined in \cite{Hong06}, when $B=1$ (slow
fading).

Finally, we note that for a specific value of $B$ and $N$  the
design of a good linear STBC is clearly impractical and a robust
solution is preferable.

% --------------------------------------------------------------
\section{Performance analysis of GST-TCM on block fading channels}
\label{sec:performance}

In this section we show the specific analysis concerning GST-TCM
\cite{Hong06}. The design of GST-TCM for slow fading ($B=1$) was
based on:
\begin{itemize}
\item the design of a trellis code that maximizes the number of non-zero
$\det(X_tX_t^\dagger)$ in (\ref{DetFL})
\item the design of partitions of the Golden code with increasing
values of $\det(X_tX_t^\dagger)$
\end{itemize}
In particular, the trellis design focused on the {\em shortest
simple error event}, i.e., a path diverging from the zero state and
remerging into the zero state in the trellis diagram. We will show
here how the length $S$ of such event influences the performance of
the code over a block fading channel.

\begin{mylemma}
A GST-TCM of length $L\geq S\geq 2$ can have $N_{s} = L-S+1$
shortest simple error events. $\hfill\blacksquare$
\end{mylemma}
{\bf Proof} -- The shortest simple error events with length $S$ can
only start in a position $\{1,2,\ldots,L-S+1\}$, thereby we obtain
$N_s = L-S+1$.$\hfill\blacksquare$

Since the codeword spans $B=L/N$ independent fading blocks of length
$N$, the simple error events will affect different blocks depending
on their starting position and length. We obtain the following
lemma.
\begin{mylemma}
A shortest simple error event of lenght $S$ is either affecting
\begin{enumerate}
\item $n_1=\lceil S/N \rceil$ consecutive blocks, or
\item $n_2=n_1+1=\lceil S/N \rceil+1$ consecutive blocks.$\hfill\blacksquare$
\end{enumerate}
\end{mylemma}
{\bf Proof} -- Depending on the starting position of the shortest
simple error event we have
\begin{itemize}
\item if $S\leq N$ then either $n_1=1$, if it is fully within one
block, or $n_2=2$.
\item if $S>N$ then it will either cross $n_1=\lceil S/N \rceil$
or $n_2=n_1+1$ concecutive blocks.
\end{itemize}

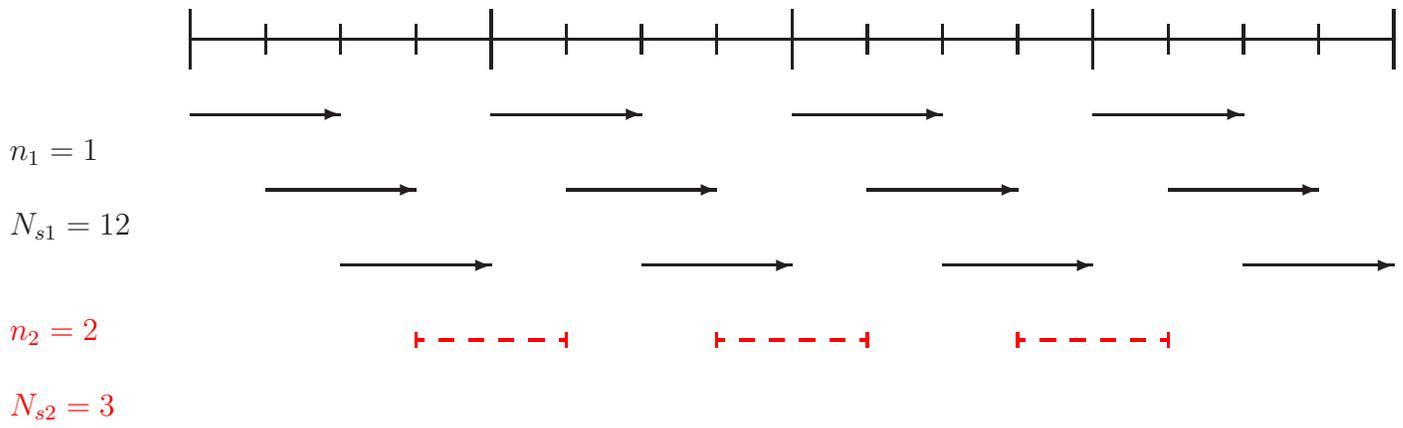
\begin{figure}[t]
\begin{center}
\setlength{\unitlength}{2mm}
\begin{picture}(90,50)(0,10) \thicklines%
\put(5,50){\line(1,0){80}} \put(5,48){\line(0,1){4}}
\put(25,48){\line(0,1){4}} \put(45,48){\line(0,1){4}}
\put(65,48){\line(0,1){4}} \put(85,48){\line(0,1){4}}
\put(10,49){\line(0,1){2}} \put(15,49){\line(0,1){2}}
\put(20,49){\line(0,1){2}} \put(30,49){\line(0,1){2}}
\put(35,49){\line(0,1){2}} \put(40,49){\line(0,1){2}}
\put(50,49){\line(0,1){2}} \put(55,49){\line(0,1){2}}
\put(60,49){\line(0,1){2}} \put(70,49){\line(0,1){2}}
\put(75,49){\line(0,1){2}} \put(80,49){\line(0,1){2}}

\put(5,45){\vector(1,0){10}} \put(10,40){\vector(1,0){10}}
\put(15,35){\vector(1,0){10}} \textcolor{red}{
\put(20,30){\dashbox(10,0)} } \put(25,45){\vector(1,0){10}}
\put(30,40){\vector(1,0){10}} \put(35,35){\vector(1,0){10}}
\textcolor{red}{ \put(40,30){\dashbox(10,0)}}
\put(45,45){\vector(1,0){10}} \put(50,40){\vector(1,0){10}}
\put(55,35){\vector(1,0){10}} \textcolor{red}{
\put(60,30){\dashbox(10,0)}} \put(65,45){\vector(1,0){10}}
\put(70,40){\vector(1,0){10}} \put(75,35){\vector(1,0){10}}

\put(-7,42){$n_1=1$} \put(-7,37){$N_{s1}=12$} \textcolor{red}{
\put(-7,30){$n_2=2$} \put(-7,25){$N_{s2}=3$} }

%\put(5,18){\vector(1,0){10}}  \put(18,17){represents a simple error
%event of length $S=2$} \put(25,10){\line(1,0){42}}
%\put(25,10){\vector(3,-2){21}} \put(46,-4){\vector(3,2){21}}
%\put(25,9){\line(0,1){2}} \put(46,9){\line(0,1){2}}
%\put(67,9){\line(0,1){2}}

%\textcolor{red}{\put(30,2){$\delta$} \put(60,2){$2\delta$}}
\end{picture}
\caption{Enumeration of simple error events of a GST-TCM with $S=2$
over a block fading channel with $B=4$ and $N=4$.}
\label{Fig:enumsee}
\end{center}
\end{figure}

For example, if $S=2$ over a block fading channel where $B=4$ and
$N=4$, as shown in Fig.~\ref{Fig:enumsee}, we have some simple error
events (solid arrows), in $n_1 = 1$ consecutive blocks and others
(dashed lines) in $n_2 = 2$ consecutive block. $\hfill\blacksquare$

\begin{mylemma}
The corresponding numbers of simple error events in Case 1 and Case
2 of the previous lemma are respectively
\begin{eqnarray}\label{NS1NS2}
 N_{s_1} = B' \times \ell ~~~~~~ N_{s_2} = N_{s}-N_{s_1}
\end{eqnarray}
where
\[
B' = B - \left\lceil \frac{S}{N} \right\rceil + 1
\]
\[
\ell = \left\lceil \frac{S}{N}\right\rceil \times N - S + 1
\]
$\hfill\blacksquare$
\end{mylemma}
{\bf Proof} -- We first recall from Lemma 2 for Case 1, that a
simple error event occupies $\lceil \frac{S}{N} \rceil$ consecutive
blocks of length $N$. Now, let us define a  {\em group} as $\lceil
\frac{S}{N} \rceil$ consecutive blocks. Hence, a group has length
$\lceil \frac{S}{N} \rceil \times N$ and contains $\ell =
\left\lceil \frac{S}{N}\right\rceil \times N - S + 1$ distinct
shortest simple error events. Since there are $B' = B - \lceil
\frac{S}{N} \rceil+ 1$ distict groups, we have $N_{s_1} = B'\times
\ell$ shortest simple error events of Case 1. The other case
directly derives from the identity $N_{s}= N_{s_1} + N_{s_2}$.
$\hfill\blacksquare$

Using the same example illustrated in Fig.~\ref{Fig:enumsee} with
$S=2$, $B=4$ and $N=4$, it is shown that we have $N_{s_1}=12$ simple
error events crossing $n_1 =1$ consecutive block (Case 1) and
$N_{s_2}=3$ simple error events crossing $n_2 =2$ consecutive blocks
(Case 2).

In order to evaluate the two dominant terms  in (\ref{eq:truncUB})
we look at the contribution of the simple error events in the
trellis together with their multiplicity. We get $N_{s_1}$ terms
with the corresponding minimum determinant
\begin{equation} \label{Deltab1}
{\Delta}^{(b)'}_1 = \min_{\ell}  \prod_{n=0}^{n_1-1} a_{\ell+n}
\end{equation}
and $N_{s_2}$ terms with the corresponding minimum determinant
\begin{equation} \label{Deltab2}
{\Delta}^{(b)'}_2 = \min_{\ell} \prod_{n=0}^{n_2-1} a_{\ell+n}
\end{equation}
%As a consequence, a better code performance can be achieved under
%the condition of a larger number $N$.

Depending on the length and structure of the simple error events,
the ${\Delta}^{(b)'}_1$ and ${\Delta}^{(b)'}_2$, together with their
multiplicity $N_{s_1}, N_{s_{2}}$, will dominate the performance of
the coding scheme.

Even if we have ${\Delta}^{(b)'}_2$ smaller than ${\Delta}^{(b)'}_1$
its contribution to the overall performance can be mitigated by the
fact that $N_{s_1} \gg N_{s_{2}}$. We will see in the following
section how the $a_\ell$s are affected by the trellis code
structure.

\section{Simulation Results}\label{simuandresults}

In this section we show the performance of different GST-TCM schemes
over block fading channels. Signal-to-noise ratio per bit is defined
as $\text{SNR}_b = n_T E_{b}/N_{0}$, where $E_{b} = E_{s}/q$ is the
energy per bit and $q$ denotes the number of information bits per
QAM symbol of energy $E_s$.

We consider two types of GST-TCM based on the two and three level
partitions $\mathbb{Z}^8/E_8$ and $\mathbb{Z}^8/L_8$ in
\cite{Hong06}. For each case we consider trellises with 4 or 16
states and 16 or 64 states, respectively. The length of the simple
error events is $S=2,3,4$ for 4,16 and 64 state trellises,
respectively. We assume the codeword length is $L=120$ and the block
fading channels are characterized by $N=1,3,5,20,40,120$. The
GST-TCM were optimized in \cite{Hong06} for the slow fading channel,
i.e., for $N=120$ (or $B=1$).

In Figures
\ref{Fig:compareTCM4stvaryblockE8}-\ref{Fig:compareTCM64stvaryblockL8}
we can see that the best performance is obtained in the slow fading
case ($N=120$), for which the codes were explicitly optimized. The
worst performance appears in the fast fading case ($N=1$), although
the difference is about 1.5-2dB at FER of $10^{-2}$ and only about
1dB at FER of $10^{-3}$. Note that the slow and fast fading curves
will eventually cross, since the fast fading exhibits a higher
diversity order. The intermediate cases of block fading exhibit a
performance between the fast and slow, which degrades as $N$
decreases.

Let us analyze these simulation results using the truncated UB
(\ref{eq:truncUB}). The sequences of values of $\det(X_t \hat{X}_t)$
in the shortest simple error events of the GST-TCMs in Figs.~
\ref{Fig:compareTCM4stvaryblockE8} to
\ref{Fig:compareTCM64stvaryblockL8} are given in Table I, where
$\delta=1/5$ is the minimum determinant of the Golden code.

Tables II-III show all the code parameters. When $N=1$ or $N=120$,
the term ${\Delta}^{(b)'}_1$ and its multiplicity $N_{s_1}$ dominate
the performance. We see that ${\Delta}^{(b)'}_1$ for $N=120$ is
always greater than that for $N=1$, provided $\delta=1/5$ and a
fixed $N_{s_1}$. This results in a better performance when $N=120$.
The same observation can be found for 64-state GST-TCM when $N=3$.

For the remaining cases, we note that ${\Delta}^{(b)'}_2$ is always
smaller than ${\Delta}^{(b)'}_1$ since $\delta=1/5$. As $N$
increases the multiplicity $N_{s_2}$ of the ${\Delta}^{(b)'}_2$ term
decreases, while $N_{s_1}$ of the ${\Delta}^{(b)'}_1$ term
increases, which results in a better performance. This analysis
qualitatively agrees with the actual performance of the codes.

\section{Conclusions}\label{conclusions}
In this letter, we analyzed the impact of a block fading channel on
the performance of GST-TCM by using a truncated UB technique. The
analysis shows that the performance of the GST-TCM designed for slow
fading channel varies slightly if the channel condition varies from
slow to fast. It is further demonstrated by simulation that the
performance degrades at most 1 dB at the FER of $10^{-3}$, when
block fading varies from slow to fast. This robust coding scheme can
be particularly beneficial for high rate transmission in WLANs using
OFDM to combat widely variable multipath fading.

%%%%%%%%%%%%%%%%%%%%%%%%%%%%%%%%%%%%%%%%%%%%%%%%%%%%%%%%%%%%%%%%%%%%
%%%%%%%%%%%%%%%%%%%%%% BIBLIOGRAPHY %%%%%%%%%%%%%%%%%%%%%%%%%%%%%%%%
%%%%%%%%%%%%%%%%%%%%%%%%%%%%%%%%%%%%%%%%%%%%%%%%%%%%%%%%%%%%%%%%%%%%
\bibliographystyle{IEEE}

\newpage\clearpage
{\bf Figures}
\begin{enumerate}
    \item Comparison of 4-state trellis codes using 16-QAM
constellation at the rate 7 bpcu form a three level partition
$\mathbb{Z}^8/E_8$ ($S=2$).
    \item Comparison of 16-state trellis codes using 16-QAM
constellation at the rate 7 bpcu form a three level partition
$\mathbb{Z}^8/E_8$ ($S=3$).
    \item Comparison of 16-state trellis codes using 16-QAM
constellation at the rate 6 bpcu form a three level partition
$\mathbb{Z}^8/L_8$ ($S=3$).
    \item Comparison of 64-state trellis codes using 16-QAM
constellation at the rate 6 bpcu form a three level partition
$\mathbb{Z}^8/L_8$ ($S=4$).
\item Enumeration of simple error events of a GST-TCM with $S=2$ over
a block fading channel with $B=4$ and $N=4$.
\end{enumerate}

\begin{figure}[t]
\begin{center}
\psfig{file=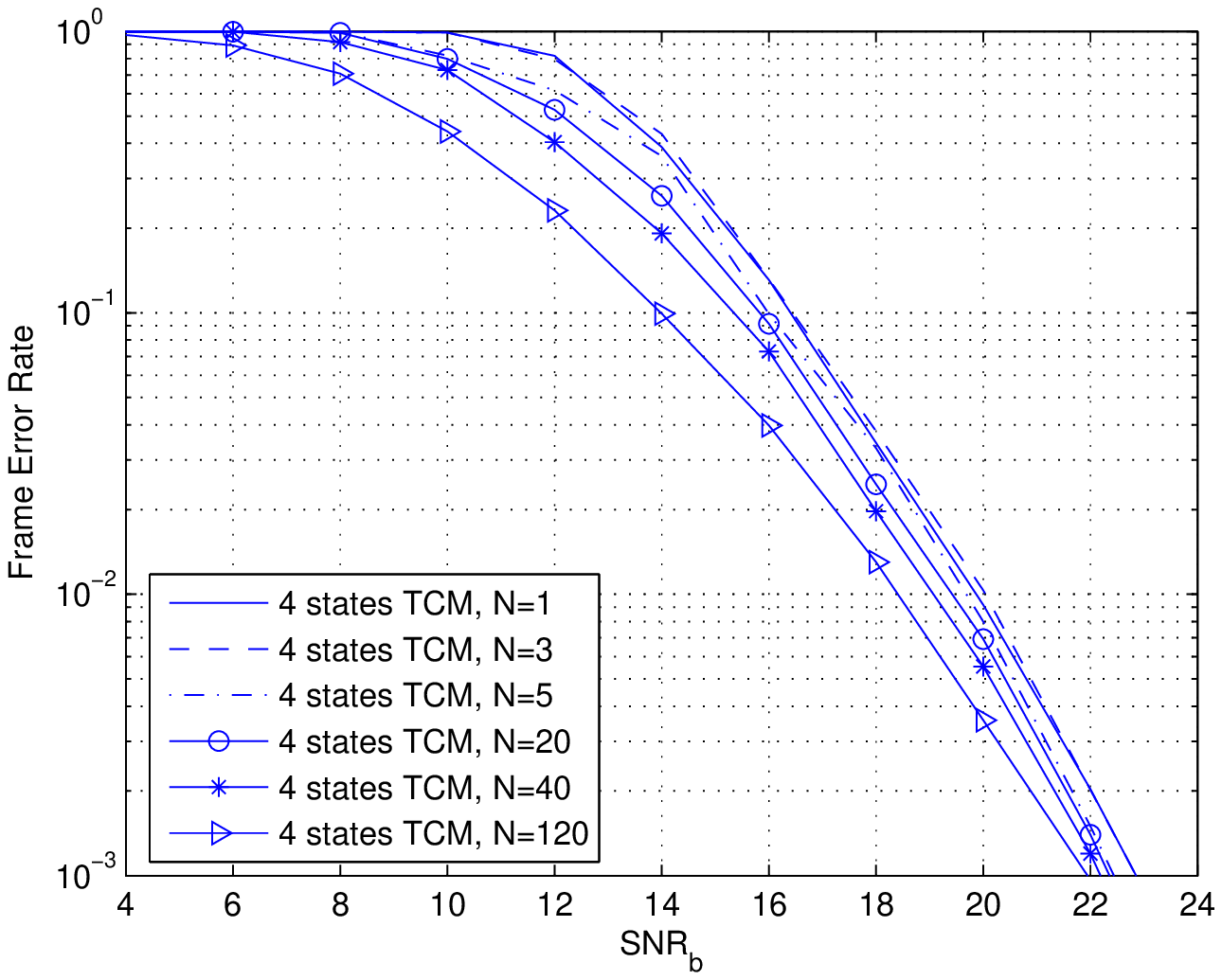,width=150mm,height=120mm}
\end{center}\vspace{-2mm}
\caption{Comparison of 4-state trellis codes using 16-QAM
constellation at the rate 7 bpcu form a three level partition
$\mathbb{Z}^8/E_8$ ($S=2$).} \label{Fig:compareTCM4stvaryblockE8}
\end{figure}

\begin{figure}[t]
\begin{center}
\psfig{file=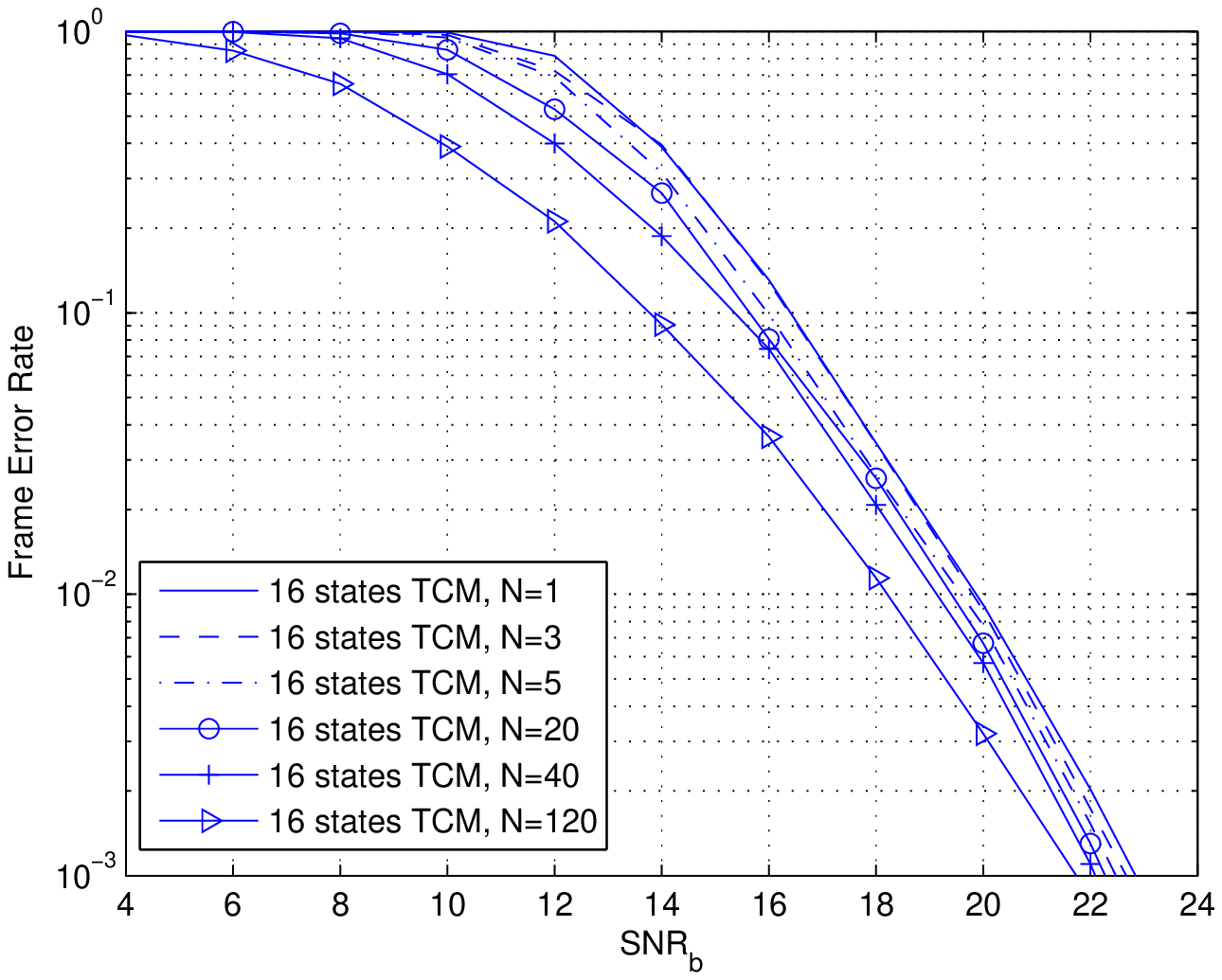,width=150mm,height=120mm}
\end{center}\vspace{-2mm}
\caption{Comparison of 16-state trellis codes using 16-QAM
constellation at the rate 7 bpcu form a three level partition
$\mathbb{Z}^8/E_8$ ($S=3$).} \label{Fig:compareTCM16stvaryblockE8}
\end{figure}

\begin{figure}[t]
\begin{center}
\psfig{file=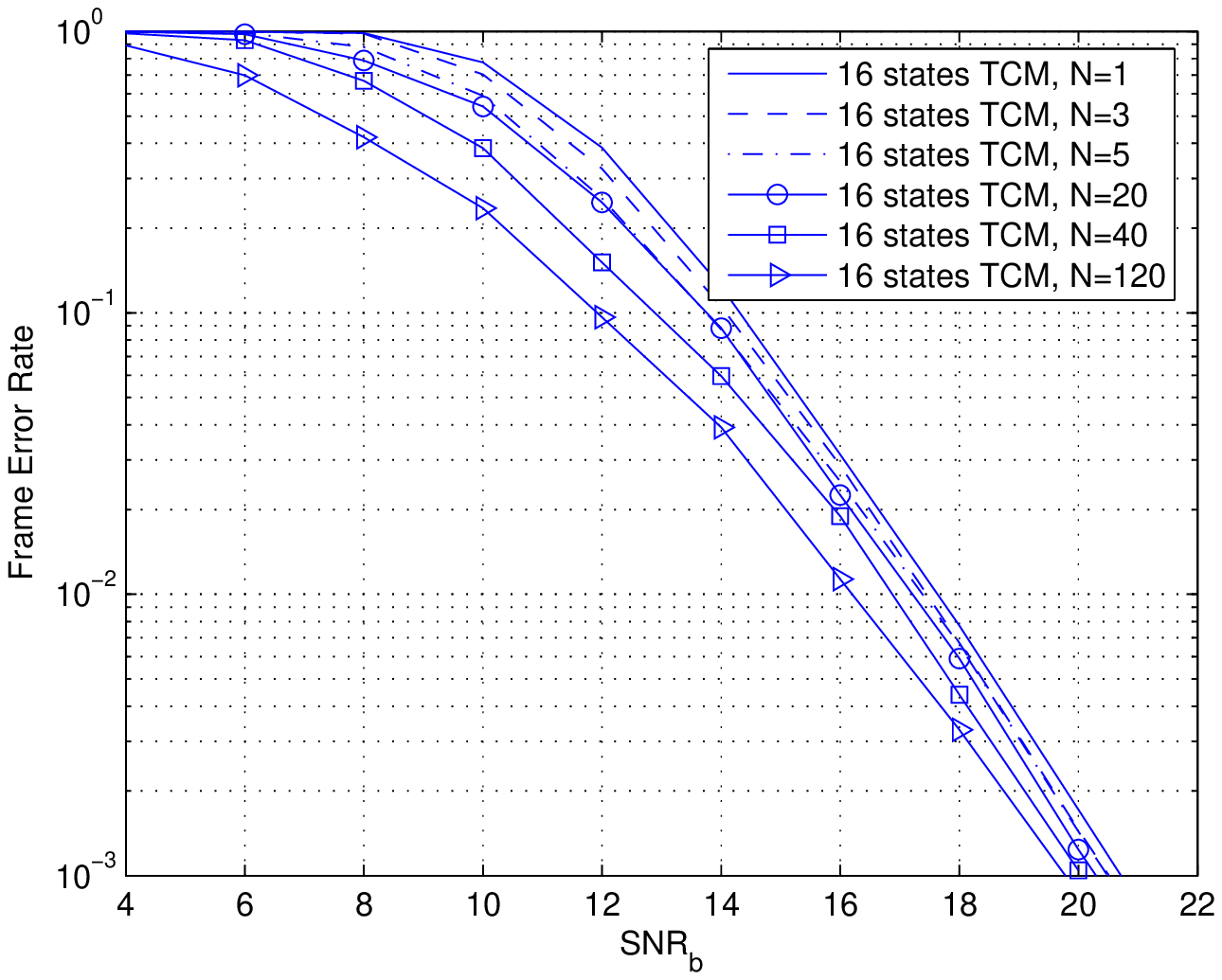,width=150mm,height=120mm}
\end{center}\vspace{-2mm}
\caption{Comparison of 16-state trellis codes using 16-QAM
constellation at the rate 6 bpcu form a three level partition
$\mathbb{Z}^8/L_8$ ($S=3$).} \label{Fig:compareTCM16stvaryblockL8}
\end{figure}

\begin{figure}[t]
\begin{center}
\psfig{file=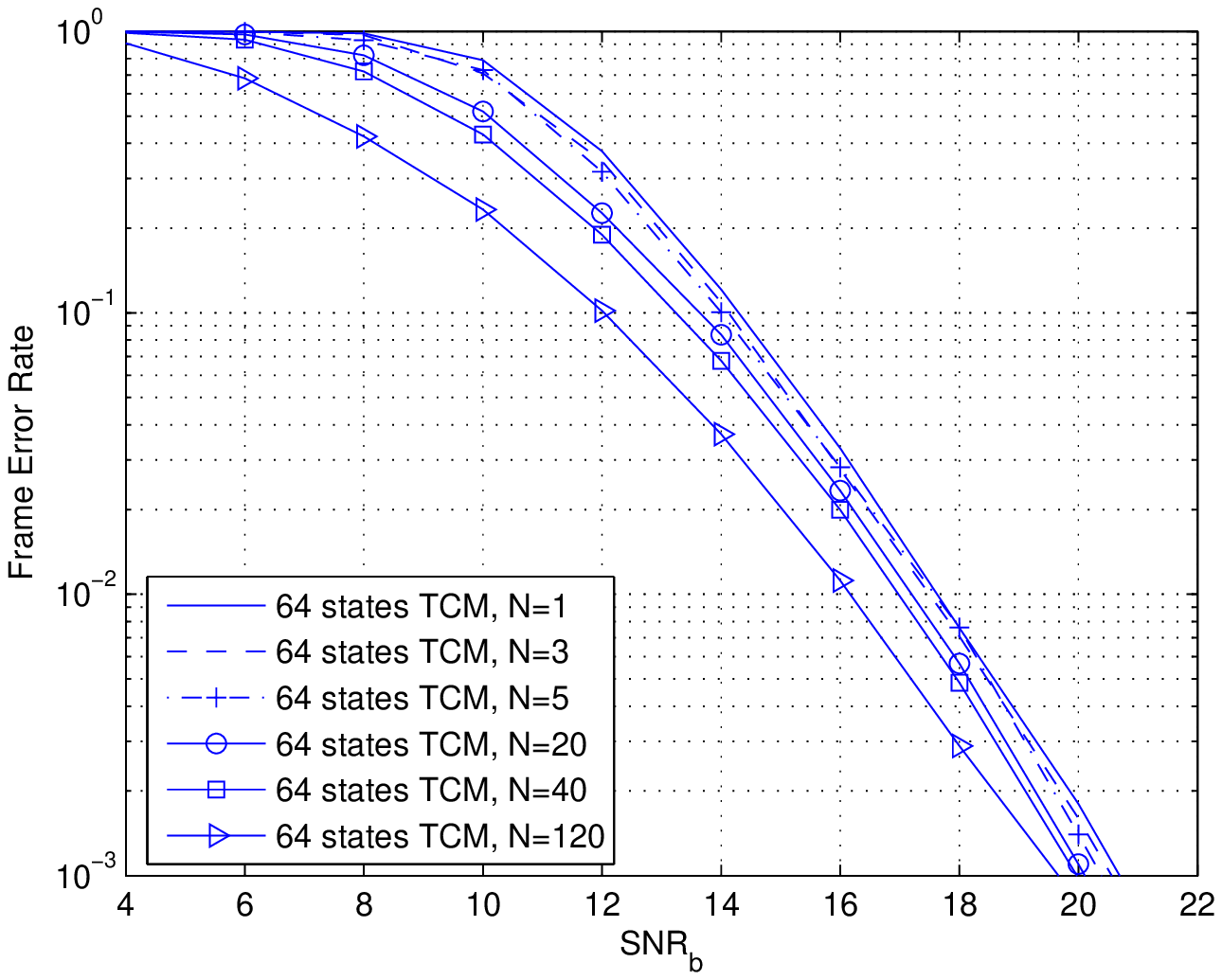,width=150mm,height=120mm}
\end{center}\vspace{-2mm}
\caption{Comparison of 64-state trellis codes using 16-QAM
constellation at the rate 6 bpcu form a three level partition
$\mathbb{Z}^8/L_8$ ($S=4$).} \label{Fig:compareTCM64stvaryblockL8}
\end{figure}

\vspace{1in} {\bf Tables}
\begin{enumerate}
    \item Sequences of $\det(X_t \hat{X}_t)$ for the
simple error events of the GST-TCMs in Figs.~
\ref{Fig:compareTCM4stvaryblockE8}-\ref{Fig:compareTCM64stvaryblockL8}
($\delta=1/5$).
    \item Simple error events for 4, 16 states $\mathbb{Z}^8/E_8$
GST-TCM, $S=2,3$ and different block fading channels
($N=1,3,5,20,40,120$).
    \item Simple error events for 16, 64 states $\mathbb{Z}^8/L_8$
GST-TCM, $S=3,4$ and different block fading channels
($N=1,3,5,20,40,120$).
\end{enumerate}
%------------------------------------------------------------------------
\newpage\clearpage
%%%%%%%%%%%%%%%%%%
%\begin{figure}[t]
%\begin{center}
%\psfig{file=Lemma1_fig.eps,width=150mm,height=140mm}
%\end{center}\vspace{-2mm}
%\caption{Enumeration of shortest simple error events with length $S$
%over $L$ Golden code alphabet symbols.} \label{Fig:Lemma1}
%\end{figure}
%%%%%%%%%%%%%%%%%%
%%%%%%%%%%%%%%%%%%%
\newpage\clearpage
%%%%%%%%%%%%%%%%%%
%%%%%%%%%%%%%%%%%%
\newpage\clearpage
%%%%%%%%%%%%%%%%%%
%%%%%%%%%%%%%%%%%%
\newpage\clearpage
%%%%%%%%%%%%%%%%%%
%%%%%%%%%%%%%%%%%%
\newpage\clearpage
%%%%%%%%%%%%%%%%%%
%%%%%%%%%%%%%%%%%%%%%%%%
\begin{table}
\begin{center}
\begin{tabular}{|c|c|c|c|c|} \hline
$S$ & step 1 & step 2 & step 3 & step 4 \\ \hline
 2 & $\delta$ & 2$\delta$ & & \\ \hline
 3 & 2$\delta$ & $\delta$ & 2$\delta$ & \\ \hline
 3 & 4$\delta$ & $\delta$ & 2$\delta$ & \\ \hline
 4 & 4$\delta$ & $\delta$ & 2$\delta$ & 4$\delta$ \\ \hline
\end{tabular}
\end{center}
 \caption{Sequences of det($X_tX_t^\dagger$) for the
simple error events of the GST-TCMs in Figs.~
\ref{Fig:compareTCM4stvaryblockE8}-\ref{Fig:compareTCM64stvaryblockL8}
($\delta=1/5$).}
\end{table}
%%%%%%%%%%%%%%%%%%%%%%%%%%%%
%%%%%%%%%%%%%%%%%%%%%%%%%%%%%%%%%%%%%%%%%%%%%%%%%%%%%%%%%%%%%%%%%%%%%%%
\begin{table}
\begin{center}
\begin{tabular}{|r|r|r|r|r|r|c|c|} \hline
\text{St.} & $N$ & $N_{s_1}$ & $ N_{s_2}$ & $n_1$ & $n_2$ &
${\Delta}^{(b)'}_1$ & ${\Delta}^{(b)'}_2$ \\ \hline \hline 4 & 1  &
119 & $-$& 2 &$-$& 2$\delta^2$ & $-$ \\ \hline 4 & 3  &   80 & 39 &
1 & 2& 3$\delta$ & 2$\delta^2$\\ \hline 4 & 5  &   96 & 23 &  1 & 2&
3$\delta$ & 2$\delta^2$\\ \hline 4 & 20 &  114 &  5 &  1 &  2&
3$\delta$ & 2$\delta^2$\\ \hline 4 & 40 &  117 &  2 &  1 &  2&
3$\delta$ & 2$\delta^2$\\ \hline 4 &120 &  119 &$-$ &  1 &$-$&
3$\delta$ & $-$\\ \hline\hline 16 &   1 &  118 &$-$ &  3 &$-$&
4$\delta^3$ & $-$ \\ \hline 16 &    3 &   40 & 78 &  1 & 2 &
5$\delta$ & 2$\delta^2+2\delta$\\ \hline 16 &    5 &   72 & 46 & 1 &
2 & 5$\delta$ & 2$\delta^2+2\delta$\\ \hline 16 &   20 &  108 & 10 &
1 & 2 & 5$\delta$ & 2$\delta^2+2\delta$\\ \hline 16 &   40 &  114 &
4 & 1 & 2 & 5$\delta$ & 2$\delta^2+2\delta$\\ \hline 16 & 120 & 118
&$-$ &  1 &$-$& 5$\delta$ & $-$ \\ \hline
\end{tabular}
\end{center} %\vspace{-3mm}
\caption{Simple error events for 4, 16 states $\mathbb{Z}^8/E_8$
GST-TCM, $S=2,3$ and different block fading channels
($N=1,3,5,20,40,120$).}
\end{table}%\vspace{-3mm}
\newpage\clearpage
%%%%%%%%%%%%%%%%%%
%%%%%%%%%%%%%%%%%%%%%%%%%%%%%%%%%%
\begin{table}
\begin{center}
\begin{tabular}{|r|r|r|r|r|r|c|c|} \hline
\text{St.} & $N$ & $N_{s_1}$ & $ N_{s_2}$ & $n_1$ & $n_2$&
${\Delta}^{(b)'}_1$ & ${\Delta}^{(b)'}_2$ \\\hline \hline   16 & 1 &
118 & $-$ & 3 &  $-$  & 8$\delta^3$ &  $-$\\ \hline 16 &   3 & 40 &
78 & 1 & 2 & 7$\delta$ & 4$\delta^2+2\delta$\\ \hline 16 &    5 & 72
& 46 & 1 & 2 & 7$\delta$ & 4$\delta^2+2\delta$\\ \hline 16 & 20 &
108 & 10 &  1 & 2 & 7$\delta$ & 4$\delta^2+2\delta$\\ \hline
16 & 40 & 114 &  4 &  1 & 2 & 7$\delta$ & 4$\delta^2+2\delta$\\
\hline 16 & 120 &  118 &   $-$  &  1 & $-$   & 7$\delta$ &  $-$\\
\hline\hline
 64 &    1 &  117 &   $-$  & 4 &  $-$ & 32$\delta^4$ & $-$ \\ \hline
 64 &    3 &  117 &   $-$  & 2 & $-$  & 28$\delta^2$, $40\delta^2$ & $-$\\ \hline
 64 &    5 &   48 & 69 & 1 & 2& 11$\delta$ & 28$\delta^2$, $40\delta^2$\\ \hline
 64 &   20 &  102 & 15 & 1 & 2& 11$\delta$ & 28$\delta^2$, $40\delta^2$\\ \hline
 64 &   40 &  111 &  6 & 1 & 2& 11$\delta$ & 28$\delta^2$, $40\delta^2$\\ \hline
  64 & 120 &  117 &   $-$  & 1 & $-$  & 11$\delta$ & $-$ \\ \hline
\end{tabular}
\end{center}%\vspace{-1mm}
\caption{Simple error events for 16, 64 states $\mathbb{Z}^8/L_8$
GST-TCM, $S=3,4$ and different block fading channels
($N=1,3,5,20,40,120$).}
\end{table}%\vspace{-3mm}
\end{document}